\documentclass[12pt]{article}
\title{\bf 
Finite Size Scaling for $O(N)$ $\phi^4$-Theory at the Upper Critical Dimension}
\author{ 
{\it R. Kenna}\\
School of Mathematical and Information Sciences,\\
Coventry University,\\
Coventry, CV1 5FB, England
}
\begin{document}
\maketitle
%-----------------------------------------------------------------------
                      {\Large
                      \begin{abstract}
%-----------------------------------------------------------------------
%
A finite size  scaling  theory for the partition function zeroes and thermodynamic
functions of  $O(N)$ $\phi^4$-theory in 
four dimensions is derived from
renormalization group methods. 
The  leading  scaling behaviour is mean-field like with
multiplicative logarithmic corrections  which  are  linked  to  the 
triviality of the theory. 
These logarithmic corrections are independent of 
$N$ for odd thermodynamic quantities and associated zeroes
and are $N$ dependent for the even ones.
Thus a numerical study of finite size
scaling in the Ising model
serves as a non-perturbative test of triviality of  $\phi^4_4$-theories 
for all  $N$.
%
%-----------------------------------------------------------------------
                        \end{abstract} }
%-----------------------------------------------------------------------
%
  \thispagestyle{empty}
%
%***********************************************************************
%
  \newpage
%
%-----------------------------------------------------------------------
                  \pagenumbering{arabic}
%-----------------------------------------------------------------------

%%%%%%%%%%%%%%%%%%%%%%%%%%%%%%%%%%%%%%%%%%%%%%%%%%%%%%%%%%%%%%%%%%%
\section{Introduction}
%%%%%%%%%%%%%%%%%%%%%%%%%%%%%%%%%%%%%%%%%%%%%%%%%%%%%%%%%%%%%%%%%%%
\setcounter{equation}{0}

The $\phi^4$-field theory, in $d$ dimensions, with an $O(N)$-symmetry 
group is of fundamental interest in both high energy and statistical 
physics. Its strongly coupled limit is the  non-linear 
$\sigma$-model, which for $N=0$ is the self-avoiding random-walk 
problem, and for  $N=1$, $2$, and $3$ is the Ising, $XY$,
and Heisenberg model, respectively. The infinite component version
is the spherical model \cite{ZJ}.

Above the upper critical dimension, $d_c=4$, 
the scaling behaviour of the $O(N)$ theory simplifies and  
the critical exponents  are  exactly those given  by mean field theory. 
It is rigorously  known  that the  continuum  limit  is  then trivial 
and described by free fields  \cite{triviality}. 
Perturbative arguments show that, for dimensionality $d>d_c$, 
the  corrections to scaling are additive in nature and are 
governed by exponents which are independent of $N$. 
The $N$-dependency of the theory resides in the amplitudes of these 
corrections \cite{Gu81}. (Notwithstanding this, the  
precise nature of finite size effects in the $O(N)$ theory
above $d_c$ has recently been the subject of some discussion 
and a better understanding is still required \cite{highd}.)

The above situation contrasts with that of low dimensional models. Indeed, 
below  the upper critical dimension
the   super-renormalizable  $\phi^4$-theory   is non-trivial 
\cite{nontriviality}.  
There, the leading critical behaviour, and even the  
existence or otherwise of a phase 
transition, is strongly dependent on the value of $N$.

At the upper  critical dimension, $d_c=4$, 
leading scaling behaviour for $\phi^4$-theory 
is coincident with mean field theory;
however it is modified by multiplicative logarithmic corrections. 
The universality class of a phase transition cannot, therefore, 
be determined through the leading 
critical exponents alone (since they are $N$-independent).
The $N$-dependency resides in the logarithmic corrections and 
these can be used, in principle,
to establish the universality class \cite{JaSt04}.

The logarithmic corrections at the upper critical dimension are intimately linked 
to the triviality of the theory there.  In fact, it has been
shown that if the leading scaling behaviour of  the susceptibility differs from
that of mean field theory only by multiplicative  logarithmic corrections 
then the $\phi^4_4$-theory must be trivial \cite{AiGr83}.  These logarithmic
corrections are the concern of this work.

The $N=4$ version of $\phi^4$-theory, in four dimensions, is of especially 
great interest to high energy physicists, as it forms the scalar sector 
of the standard model where it plays a central r\^ole in  the  generation of 
fermion and gauge boson mass via the Higgs mechanism. 
Triviality of the $O(4)$
$\phi^4$-theory in four dimensions is therefore an especially important issue.
There is an abundance of analytical 
\cite{ZJ,WeRi73,BLZ,HaTa87,Kl01} and numerical 
\cite{ADCCaFr,RuGu85,LW,LaMo90,KeLa93,La93,AdSt97,BuCo00,N0logs,Ak01,BiMa02} 
 evidence in favour of
 triviality, mostly for the single-component theory. 
However, since a complete rigorous 
proof is still lacking, various ideas aimed at recovering a non-trivial
continuum theory are also being explored \cite{CeCo04}.

In high energy physics, lattice gauge theories are also described by a mean field
equation of state with logarithmic corrections.
The issue of logarithmic triviality of these and related
models in four dimensions was  explored in  \cite{QEDtriv}.
Multiplicative logarithmic corrections in four dimensions also play an important role in 
non-equilibrium phase transitions (and in directed percolation  in particular 
\cite{JaSt03,LuWi04}).
Besides physics, such systems are of direct interest in biology and chemistry 
\cite{JaSt04,Ja81}
as well as in the study of avalanches in sandpile models \cite{BaTa88,Lu98}.

The renormalization group yields predictions on how criticality is approached 
in the infinite-volume four-dimensional case \cite{WeRi73,BLZ,LW}. 
With the reduced temperature, $t$, denoting the distance from criticality,
the  perturbative formula for the susceptibility is
\begin{equation}
 \chi_\infty (t)  
 \propto 
 |t|^{-1}
 (-\ln{|t|})^{\frac{N+2}{N+8}}
 \left\{ 
         1 + O \left(
                      \frac{
                             \ln{ (-\ln{|t|})} 
                            }
                             {
                              \ln{|t|}
                            }
               \right)
 \right\}
\quad ,      
\label{two}
\end{equation}
while that of the singular part of the specific heat is
\begin{equation}
  c_\infty (t)  =  
  \left\{ 
          \begin{array}{ll}
                            (-\ln{|t|})^{\frac{4-N}{N+8}} 
        \left\{ 1 + O \left(
                             \frac{
                                   \ln{ (-\ln{|t|}) }
                                  }
                                  {
                                   \ln{|t|}
                                  }
                       \right)
         \right\}
                                                    &  {\rm if\ }  N \neq  4
                        \\
                         \ln{(-\ln{|t|})} 
        \left\{ 1 + O \left(
                       \frac{
                             1 
                            }
                             {
                              \ln{|t|}
                            }
                      \right)
 \right\}
                                                     & {\rm if \ }  N = 4         
                     \quad ,
          \end{array}
  \right. 
\label{three}   
\end{equation}
and that of the correlation length  is 
\begin{equation}
 \xi_\infty (t) = |t|^{-\frac{1}{2}}
 \left(
  - \ln{|t|}
 \right)^{\frac{N+2}{2(N+8)}}
\left\{ 1 + O \left(
                      \frac{
                             \ln{ (-\ln{|t|})} 
                            }
                             {
                              \ln{|t|}
                            }
               \right)
 \right\}
\quad .
\label{four}
\end{equation} 
(In (\ref{two}), and henceforth in this paper, $\chi$ refers to the longitudinal 
susceptibility in the case where  $N \ge 2$ and $t<0$.)

In each of these formulae, the leading power-law scaling behaviour 
coincides with the predictions of mean field  theory. One notes that
the multiplicative logarithmic corrections are $N$-dependent 
in each case.
Equations (\ref{two})  and (\ref{four}) have been rigorously  proven for the 
weakly coupled version of the single-component theory \cite{HaTa87}.
No proof exists for the strongly coupled or general-$N$ theories.

Finite size scaling [FSS] is an important tool used in the determination
of universality classes 
and most modern Monte Carlo studies expoit it to determine critical exponents \cite{FSS}.
In particular, recent analyses of systems at the upper
critical dimension have adopted a finite-sized approach
\cite{Ak01,Lu98,LuBl97,BaFe98,Ru98,CaGa01,AkEr01,RoLu02}. 

FSS was formulated as a hypothesis by Fisher and co-workers \cite{FSS} and 
proved for $d<d_c=4$ on the basis of the renormalization group  by Br\'ezin \cite{Br82}.
Let $P_L(t)$ be a thermodynamic quantity measured on a system of linear extent $L$
at a distance $t$ from the bulk critical point.
The traditional FSS hypothesis states that if $P_\infty$ exhibits 
an algebraic singularity (at $t=0$ in the infinite system), then its finite size behaviour
(in this region) is given by 
\begin{equation}
  \frac{P_L(t)}{P_\infty(t)}
  =
  {\cal{F}}_P\left(
   \frac{L}{\xi_\infty (t)}
  \right)     
\quad ,
\label{FSShyp}
\end{equation}
where $\xi_\infty (t)$ 
is the correlation length of the infinite-size system. 
Here,   ${\cal{F}}_P$  is  an a priori
unknown function of its argument which is called the scaling variable.

Br\'ezin's theoretical justification of (\ref{FSShyp})
relies on two assumptions over   and above the usual assumptions of renormalization group
 theory. These 
are  that  {\em{(i)}} the  system  length,  $L$, is  not  renormalized  in the flow 
equations  and  that {\em{(ii)}} the infra-red fixed point  is not 
zero. This  latter  assumption fails in and above four dimensions.

At the upper critical dimension, then, the FSS formulae have had to be calculated 
directly from the (perturbative) renormalization group equations \cite{Br82}. 
The $N=0$ case introduces significant simplifications which were exploited in \cite{ADCCaFr} 
to test the renormalization group predictions for logarithmic corrections 
 numerically. (However precise agreement in this case is obscured by large 
subleading corrections \cite{N0logs}.)

Early attempts to verify the perturbative predictions numerically, in the Ising case,
were hampered by limited computational   resources \cite{RuGu85}. 
Logarithmic triviality of the Ising model in four dimensions was eventually 
established using sophisticated techniques in \cite{KeLa93}.
At present, no direct calculations have been carried out for finite $N > 1$. 
Given the recent enormous advances in computer hardware and algorithms, direct
computational detection of logarithms in high dimensional $O(N)$ systems is now
feasible and it is reasonable to ask what results might be expected.

The purpose of this paper is twofold. Firstly, some  results 
concerning the FSS of the $O(N)$ model are presented, paying particular attention to the 
multiplicative logarithmic corrections. 
Wherever possible, these results are compared with ones already available in the literature
(specifically for the Ising and spherical models)
and  in each case complete agreement is found. 
It turns out that the leading logarithmic corrections to the finite size dependency 
are independent
of $N$ in the odd sector. This means that numerical confirmation of their
existence in the Ising model also serves as a non-perturbative test for all $N$ 
\cite{KeLa93}. 
Secondly, all of the renormalization group
 finite size results can be recovered from a modified version of the
FSS hypothesis, extending its validity to the upper critical dimension.

The layout of the remainder of this paper is as follows.
For completeness, renormalization group techniques are recalled in Sec.\ref{sec:rg}
and applied to the finite-sized $O(N)$ four-dimensional theory.
The substantive results concerning the detailed $N$-dependence of the logaritmic
corrections to FSS for the partition function zeroes and thermodynamic functions
are presented in Sec.\ref{sc:pfz}.
The general applicability of the modified FSS hypothesis, valid
at and below the upper critical dimension, is verified in Sec.\ref{sc:FSS}
and conclusions are drawn in Sec.\ref{sc:ccl}.

%%%%%%%%%%%%%%%%%%%%%%%%%%%%%%%%%%%%%%%%%%%%%%%%%%%%%%%%%%%%%%%%%%%
\section{Finite Size Renormalisation Group Analysis}
%%%%%%%%%%%%%%%%%%%%%%%%%%%%%%%%%%%%%%%%%%%%%%%%%%%%%%%%%%%%%%%%%%%
\label{sec:rg}
\setcounter{equation}{0}

The starting point is the   action   for  the  $N$-component  $\phi^4$-theory  
in  $d$-dimensional space, which is
\begin{equation}
 S
 = 
 \int{ d^dx
 \left[
  \frac{1}{2}(\nabla\mbox{\boldmath{$ \phi$}})^2 
        + \frac{m_0^2}{2}{\mbox{\boldmath{$\phi$}}}^2 
            + \frac{g_0}{4!}(\nabla{\mbox{\boldmath{$ \phi$}}}^2)^2 + 
              {\mathbf{H}}(x)\nabla{\mbox{\boldmath{$ \phi$}}}(x)
 \right]
}
\quad ,
\label{action}
\end{equation}                                                         
where  $\mbox{\boldmath{$\phi$}}$ is an  $N$-dimensional vector,
$m_0$   the bare bosonic mass
in the quantum field theory,
$g_0$  the bare quartic self-coupling and 
${\mathbf{H}}$ an external field.   Writing ${m_0}^2 = {m_0}_c^2 + t(x)$, 
where ${m_0}_c$ is the critical bare    mass for which the renormalized 
theory is massless,  $t(x)$ becomes  a source for quadratic composite fields. Its inclusion
facilitates the derivation of energy-energy correlation functions (i.e., the 
specific heat). If $t$ is allowed to become independent of $x$, it is
a measure   of the distance        from 
criticality \cite{BLZ}.       
In the statistical physics theory,   the squared mass 
corresponds to the Boltzmann factor and $t$ is called       the reduced 
temperature.

The generating functional for the connected Green's functions for the $N$-component 
theory  is
\begin{equation}
  \exp{W[t,{\mathbf{H}}]}
  \propto
  \int{\prod_x{ \prod_\alpha{ d \phi_\alpha (x) 
  \exp{\left( -S \right)}}}}
\quad ,         
\label{thesis2.7}
\end{equation}
where the constant of proportionality  is such that $W[0,{\bf{0}}] = 0$,
and $\alpha = 1,\dots,N$ label the field components. 
The function conjugate to $H_\alpha$ is 
\begin{equation}
 M_\alpha (t,{\mathbf{H}},x) = \frac{\delta W[t,{\mathbf{H}}]}{\delta H_\alpha(x)} 
\label{thesis2.9}
\end{equation}     
and facilitates the following expression for the generating functional  
for one-particle irreducible vertex  
functions:
\begin{equation}
\Gamma [t,{\mathbf{M}}] + W[t,{\mathbf{H}}] 
 = \int{dx {\mathbf{H}}(x){\mathbf{M}}(t,{\mathbf{H}},x)}
\quad .
\label{thesis2.10}
\end{equation}
This gives 
\begin{equation}
 H_\alpha(x)
 = 
 \frac{\delta \Gamma [t,{\mathbf{M}}]}{\delta M_\alpha }
\quad .
\label{thesis2.11}
\end{equation}
    
The four-dimensional version  of  the  theory  posesses certain 
subtleties  not  present  in lower  dimensions.  The   infra-red fixed point of the  
Callan-Symanzik function  moves to the origin as  the dimensionality    
becomes  four  (i.e., the  perturbative  approach  predicts that the theory is 
trivial) and  becomes a double zero. This latter phenomenon
is responsible for  the  occurrence  
of   multiplicative logarithmic   corrections.   
There is also an 
inhomogeneous term in the  renormalization group equations. 
The graph responsible for this   term is 
not    divergent in less than   four dimensions where singular behaviour 
comes from the homogeneous term. In four dimensions  the   inhomogeneous 
term can and does contribute to the leading scaling    behaviour.  

The extension from the single 
component  $\phi_4^4$-theory   
to the $N$-component    version is straightforward \cite{BLZ,RuGu85,KeLa93}. 
The critical theory is firstly renormalized.      The bulk renormalization  constants are 
sufficient   to   renormalize    the   finite size   theory and this 
renormalization is performed at an arbitrary non-zero mass parameter, 
$\mu$, in order to control infra-red divergences.  One then applies a Taylor expansion 
about $t=0$ and ${\mathbf{M}}={\mathbf{0}}$  to the vertex functions to find 
the renormalization group equations for the    massive   theory. 
So even though the symmetry might be (explicitly or spontaneously) 
broken, the vertex functions   in   the   critical   region are
evaluated with ${\mathbf{M}}={\mathbf{0}}$.  The question of whether 
there exists spontaneous breaking of symmetry (in the infinite-volume case) 
is answered by appealing 
to (\ref{thesis2.11}) with ${\mathbf{H}} = {\mathbf{0}}$ and searching for solutions 
for which ${\mathbf{M}} \ne {\mathbf{0}}$ \cite{Amit}.

Following 
\cite{BLZ,RuGu85,KeLa93} and using dimensional analysis, the solution for 
the renormalized free energy is given by
\begin{equation}
       f_R(t,{\mathbf{M}},g_R,\mu,L)   
 =
       L^{-4}
    f_R(L^2t(\lambda),L{\mathbf{M}}(\lambda),
        g_R(\lambda),L\mu(\lambda),1)  
       + \Pi(\lambda;t) 
\quad  ,
\label{thesis5.2}
\end{equation}      
where
\begin{equation}
 \Pi(\lambda;t) =       -  
       \frac{1}{2!}
       \int_1^\lambda 
           \frac{d\lambda'}{\lambda'} 
           {t(\lambda')}^2
           \Upsilon(g_R(\lambda^\prime))
\quad .
\end{equation}        
Here, $\mu (\lambda) = \lambda \mu$ is a rescaling of the arbitrary mass
$\mu$. The functions $g_R(\lambda)$, ${\mathbf{M}} (\lambda)$,   $t(\lambda)$ 
and $\Upsilon (g_R(\lambda))$ respond to this rescaling through the      flow 
equations.  To leading order in perturbation theory the flow equations
for  the $O(N)$ theory are \cite{BLZ,RuGu85}
\begin{eqnarray}
  \frac{ d g_R (\lambda) }{ d \ln{\lambda} } 
  & = & 
  \frac{N+8}{6} 
  {g_R(\lambda)}^2  
  \left\{ 1+ O(g_R ( \lambda )) \right\}
\quad ,
\label{thesis5.69}
\\
 \frac{d \ln{t(\lambda)}}{d \ln{\lambda}} 
  & = &  \frac{N+2}{6} g_R(\lambda)
   \left\{ 1+ O({g_R (\lambda)}^2) \right\}
\quad ,
\label{thesis5.71}
\\
  \frac{d \ln{M_\alpha}(\lambda)}{d \ln{\lambda}} 
 & = & - \frac{N+2}{144} g_R(\lambda)^2   
    \left\{ 1+ O(g_R (\lambda)) \right\}
\quad ,
\label{thesis5.70}  
\\
  \Upsilon (g_R(\lambda)) & = & \frac{N}{2}\left\{ 1
        + O(g_R (\lambda)) \right\}
\quad .
\label{thesis5.72}
\end{eqnarray}                 
For $\lambda \ll 1$ the solutions to these  equations are
\begin{eqnarray}
 g_R(\lambda ) & = & a_2 (-\ln{\lambda})^{-1}    
\quad ,
\label{g1}
\\
 t (\lambda )  & = & a_1  t (-\ln{\lambda})^{-\frac{N+2}{N+8}}
\quad ,
\label{t1}
\\
 M_\alpha (\lambda ) & = & b_1 M_\alpha 
\label{M1}
\quad ,
\end{eqnarray}
\begin{equation}
  \Pi(\lambda;t) \propto   \left\{ 
                           \begin{array}{ll}
  a_3 t^2 
              (-\ln{\lambda})^{\frac{4-N}{N+8}}
              &  {\rm for\ } N \neq 4 
                                                                \\
     b_2   t^2 
                \ln{(-\ln{\lambda})} 
                         & {\rm for\ }  N = 4          
               \quad ,
               \end{array}
          \right. 
\label{thesis5.84'}
\end{equation}
where  the $a_j$ have the form
\begin{equation}
  a \left\{ 1 + 
        O\left( \frac{\ln{|\ln{\lambda}|}}{\ln{\lambda}}
  \right)  \right\}
\quad ,
\label{1stc}
\end{equation}
and the $b_j$ are
\begin{equation}
 b
 \left\{ 
          1 + O\left( 
                     \frac{1}{\ln{\lambda}}
               \right)
 \right\}     
\quad .                           
\label{2ndc}
\end{equation}
The $a_j$ and $b_j$ are dependent on $N$ through the prefactors of 
(\ref{1stc}) and (\ref{2ndc})  and 
there is no other $N$ dependency in  (\ref{g1}) or (\ref{M1}).
This is due to 
the fact that, in perturbation theory  in four dimensions,
(\ref{thesis5.69})  and (\ref{thesis5.70})  
begin with quadratic terms in the running coupling.
Only the logarithmic term coupled to  the reduced temperature $t$ in (\ref{t1}) has an 
$N$-dependent exponent 
and this is due to the term linear in $g_R(\lambda)$ in (\ref{thesis5.71}). 

Now, choosing $\lambda = L^{-1}$ and applying perturbation theory to the 
homogeneous  term  of (\ref{thesis5.2}), one finds
\begin{equation}
         f_R\left( t,{\mathbf{M}},g_R,1,L  \right)
 = a_1^\prime t M^2 \left( \ln{L}  \right)^{-\frac{N+2}{N+8}}
   +
   a_2^\prime  M^4 (\ln{L})^{-1}
   +
   \Pi (L^{-1};t)
\quad ,
\label{thesis5.99}
\end{equation}
where $M = |{\mathbf{M}}|$ and  $a_1^\prime$ and $a_2^\prime$ are of the form (\ref{1stc})   
above,
with $\lambda$ replaced by $L^{-1}$.

Applying (\ref{thesis2.11}) to this 
yields, for the ($x$-independent) external field,
\begin{equation} 
  H_\alpha \left( t,{\mathbf{M}},g_R,1,L  \right)
 =
 2 a_1^\prime t M_\alpha (\ln{L})^{-\frac{N+2}{N+8}}
 + 
 4 a_2^\prime {M}^2 M_\alpha (\ln{L})^{-1}
\quad .
\label{thesis5.101} 
\end{equation}
Note, again, that there is no $N$ dependence in the logarithm in the second term 
on the right hand side of (\ref{thesis5.99}) and consequently in the logarithmic 
part of 
the last term of (\ref{thesis5.101}). 
This crucial fact will ultimately lead to the $N$ independence of the 
multiplicative logarithmic corrections to the FSS of the Lee-Yang
zeroes and associated odd thermodynamic functions. 

Indeed, $N$ independence 
in the odd sector can already be seen in the infinite-volume magnetization.
There, the expressions for $f_R$ and
$H = |{\mathbf{H}}|$ are similar to those of (\ref{thesis5.99}) and  (\ref{thesis5.101}),
except that $\ln{L}$ is replaced by $\ln{M}$ \cite{BLZ}. The strategy 
there is to set $H=0$ and to eliminate the spontaneous magnetization 
in favour of $t$. (Thus the thermodynamic 
scaling behaviour (scaling with $t$) in (\ref{two}) to (\ref{four}) exhibit 
$N$-dependent logarithmic corrections.) If, however, $t$ is set to zero
in the infinite-volume counterpart of (\ref{thesis5.101}), one 
finds, up to leading logarithms (using $H_\alpha = H \delta_{\alpha,1}$ to isolate the 
first component as the longitudinal one),
\begin{equation}
 M(H) \propto H^{\frac{1}{3}} (-\ln{H})^{\frac{1}{3}}
\quad ,
\label{empty}
\end{equation}
independent of $N$ \cite{BLZ}.

From the Legendre transformation (\ref{thesis2.10}), and from 
(\ref{thesis5.99}) and (\ref{thesis5.101}), the free energy per 
unit volume in the presence of an external field  is
\begin{equation}
 W_L(t,{\bf{H}})
 = a_1^\prime  t {{M}}^2 (\ln{L})^{-\frac{N+2}{N+8}}  
   +
   3 a_2^\prime  {{M}}^4   (\ln{L})^{-1}
   -
   \Pi(L^{-1};t)
\quad .
\label{thesis5.103}
\end{equation}     
From this expression  the FSS relations can be derived.
From (\ref{thesis5.101}) and (\ref{thesis5.103}), if $t=0$,
\begin{equation}
 W_L(0,{\mathbf{H}})
 \propto
 H^{\frac{4}{3}}
 (\ln{L})^{\frac{1}{3}}
\left\{ 1 + O\left( \frac{\ln{(\ln{L})}}{\ln{L}}
  \right)  \right\}
\quad .
\label{thesis7.21}
\end{equation} 
On the other hand, when the external magnetic field vanishes, 
(\ref{thesis5.101}) gives $M=0$ (corresponding to the symmetric 
phase) or (up to additive correcion terms)
\begin{equation}
 {{M}}^2 \propto (-t) (\ln{L})^{\frac{6}{N+8}}
\left\{ 1 + O\left( \frac{\ln{(\ln{L})}}{\ln{L}}
  \right)  \right\}
\quad ,
\end{equation}
which corresponds to the $t<0$ phase. In either case,
(\ref{thesis5.103}) gives
\begin{equation}
  W_L(t,{\mathbf{0}})   
  \propto   
  \left\{ 
         \begin{array}{ll}
          t^2 
                (\ln{L})^{\frac{4-N}{N+8}}
                       \left\{ 1 + O\left( \frac{\ln{(\ln{L})}}{\ln{L}}
                            \right)  \right\}
                                         &  {\rm if\ }  n \neq  4
                                                               \\
          t^2
                \ln{(\ln{L})}
\left\{ 1 + O\left( \frac{1}{\ln{L}}
  \right)  \right\}
                                         & {\rm if \ }  n = 4         
\quad .
               \end{array}
          \right. 
\label{thesis7.16}   
\end{equation}
This functional form for the free energy holds regardless of phase.

%%%%%%%%%%%%%%%%%%%%%%%%%%%%%%%%%%%%%%%%%%%%%%%%%%%%%%%%%%%%%%%%%%%
\section{Partition Function Zeroes and Thermodynamic Functions}
%%%%%%%%%%%%%%%%%%%%%%%%%%%%%%%%%%%%%%%%%%%%%%%%%%%%%%%%%%%%%%%%%%%
\setcounter{equation}{0}
\label{sc:pfz}

The zeroes of the partition function present an analytically powerful
and numerically precise approach to the study of critical phenomena,
which is complimentary to the more traditional functional approach.
Zeroes in the complex external-field plane are referred to as Lee-Yang
zeroes \cite{LY}  while their complex-temperature counterparts are Fisher 
zeroes \cite{Fi64}. In particular, the FSS behaviour of these zeroes
can be used to extract ratios of critical exponents. The precise 
scaling formulae were given for $d<d_c$ in \cite{IPZ} and for the single-component
four-dimensional $\phi^4$-theory in \cite{KeLa93}. The $O(N)$-generalization 
of the latter can now be derived from the expressions 
(\ref{thesis7.21}) and (\ref{thesis7.16})   for the free energy.

From the expression (\ref{thesis7.21}), for the finite-size free energy 
at $t=0$, the partition function must take the form
\begin{equation} 
 Z_L(0,H) 
 = 
 Q\left(
 H^{\frac{4}{3}} L^4 (\ln{L})^{\frac{1}{3}} 
 \left\{ 1 + O\left( \frac{\ln{(\ln{L})}}{\ln{L}}
  \right)  \right\}
 \right)
\quad ,
\label{tbi}
\end{equation}
for  some  unknown  function $Q$.   At  a complex Lee-Yang zero, $H_j(L)$, this 
vanishes.  Following \cite{IPZ}, and inverting (\ref{tbi}),
 the FSS formula for the  Lee-Yang zeroes is found 
to be
\begin{equation}
 H_{j}(L) 
 \sim
 L^{-3}
 (\ln{L})^{-\frac{1}{4}}
 \left\{
      1 + O\left(
                  \frac{\ln{(\ln{L})}}{\ln{L}}
           \right)
 \right\}
\quad .
\label{FSSLY}
\end{equation}
To reiterate, there 
is no $N$ dependence in the power of the logarithmic corrections here for the following reason.
The expression (\ref{FSSLY}) is a direct consquence of (\ref{thesis7.21}), which itself
comes from the free energy in (\ref{thesis5.103}). There, only the logarithmic terms 
coupled to the reduced temperature, $t$, are $N$-dependent, for the reasons explained 
in Sec.~\ref{sec:rg}. The $N$ independency of the remaining terms is due to the
fact that the perturbative expressions
 (\ref{thesis5.69}) and (\ref{thesis5.70}) 
begin with quadratic terms in the running coupling.
In the derivation of the FSS behaviour of the Lee-Yang zeroes,
the reduced temperature, $t$,  has been set to zero and the associated
$N$-dependent logarithms of (\ref{thesis5.103}) are therefore absent.
Consequently, there will also 
be no $N$ dependency in the multiplicative
logarithmic corrections to derivable thermodynamic functions such as the magnetic susceptibility.

Writing   the 
partition function as a product over its Lee-Yang zeroes,
\begin{equation}
 Z_L(t,H) \propto \prod_j{\left( H-H_j(L)  \right)}
\label{old4.9}
\quad ,
\end{equation}
the (longitudinal) 
magnetic susceptibility  is directly derived as  the second derivative of the free 
energy with respect to $H$, 
\begin{equation}
 \chi_L(t,H) \propto \frac{1}{L^4}
 \sum_j{
        \frac{1}{\left(H-H_j(L) \right)^2}
       }
\quad .
\end{equation}
The zero field susceptibility is then
\begin{equation}
 \chi_L \left( t \right) 
 \propto \frac{1}{L^4} \sum_j{ \frac{1}{H_j(L)^2} } 
\quad .
\label{fsr}
\end{equation}
It is reasonable (and usual \cite{KeLa93,XY}) 
to assume that the scaling of the susceptibility is dominated by the 
behaviour of the zeroes closest to the real axis. Then, (\ref{FSSLY}) 
and (\ref{fsr}) give the FSS formula
\begin{equation}
 \chi_L(0) \propto L^2 (\ln{L})^{\frac{1}{2}}
 \left\{
        1 + O\left( 
                   \frac{\ln{(\ln{L})}}{\ln{L}}
             \right) 
 \right\}
\quad .
\label{fssforchiin4d}
\end{equation}            
This is independent of $N$, as anticipated.

Similar reasoning can be used to determine the FSS behaviour of the
Fisher zeroes using (\ref{thesis7.16}).
Setting  the 
corresponding partition function to zero and solving  for   $t$  gives 
the following FSS formula for the Fisher zeroes  in  four  dimensions;
\begin{equation}
 t_j(L)
 \sim   
 \left\{ 
  \begin{array}{ll}
                     L^{-2} (\ln{L})^{\frac{N-4}{2(N+8)}}\left\{
        1 + O\left( 
                   \frac{\ln{(\ln{L})}}{\ln{L}}
             \right) 
 \right\}

                                         &  {\rm if\ }  n \neq  4        
                                                              \\
                     L^{-2} (\ln{(\ln{L})})^{-\frac{1}{2}} \left\{
        1 + O\left( 
                   \frac{1}{\ln{L}}
             \right) 
 \right\}

                                         & {\rm if \ }  n = 4          
  \quad .
  \end{array}
 \right. 
\label{FSSF}
\end{equation} 

From the expression (\ref{FSSF}), the FSS of the even thermodynamic functions are easily derived.
In the absence of an external ordering field, the partition function may be written
\begin{equation}
 Z_L(t,0) \propto \prod_j{\left( t-t_{j}(L) \right)}
\quad .
\end{equation}
The specific heat, given by the second derivative of the free energy
with respect to $t$, is 
\begin{equation}
 C_L(t) = -\frac{1}{L^4}
 \sum_j{\frac{1}{\left(t-t_{j}\right)^2}}
\quad .
\end{equation}            
At the bulk critical point, $t=0$, this becomes
\begin{equation}
 C_L \left( 0 \right) 
 = -\frac{1}{L^4} \sum_j{ \frac{1}{t_{j}(L)^2} }
\quad ,
\end{equation}
which, together with (\ref{FSSF}) and the reasonable assumption
that the first few zeroes dominate scaling \cite{KeLa93,XY},
 yields
\begin{equation}
 C_l(0)  
 \sim
 \left\{ 
 \begin{array}{ll}
          (\ln{L})^{\frac{4-N}{N+8}}
\left\{
        1 + O\left( 
                   \frac{\ln{(\ln{L})}}{\ln{L}}
             \right) 
 \right\} 
                                         &  {\rm if\ } n \neq 4
                                                             \\
         \ln{(\ln{L})} 
\left\{
        1 + O\left( 
                   \frac{1}{\ln{L}}
             \right) 
 \right\}
                                        &  {\rm if\ } n = 4          
 \quad .
 \end{array}
 \right. 
\label{FSSCv}   
\end{equation}
In particular, the $N=1$ result recovers the theoretical results of 
\cite{RuGu85} and \cite{LaMo90}
for the singular part of the Ising specific heat  and 
the $N \rightarrow \infty$ limit recovers the result  of
\cite{SiPa92} for the spherical model in four dimensions. 
The result (\ref{FSSCv}) is also in agreement with a {\emph{conjecture}}
made in \cite{SiPa92}, except in the case of $N=4$. In that case, 
$\log$-$\log$ corrections are dominent in (\ref{FSSCv}).
Furthermore, the $N$-independent result (\ref{fssforchiin4d})
is coincident with the Ising result of \cite{LaMo90}
as well as with results for the spherical model
(given by $N \rightarrow \infty$) explicitly obtained
in \cite{SiPa92} and seperately in \cite{ShRu86}.

The real part of the first Fisher zero may be considered a pseudocritical point.
For general $N \ne 4$, 
(\ref{FSSF}) gives that this pseudocritical point 
approaches the critical one as $L^{-2}(\ln{L})^{(N-4)/2(N+8)}$.
In the $N=1$ case,
this is coincident with the specific-heat pseudocritical-point scaling,
$L^{-2}(\ln{L})^{-1/6}$,
derived from direct renormalization group considerations of the Ising
model in \cite{RuGu85} and with the equivalent susceptibility peak position
derived in \cite{LaMo90}. For $N=4$ the pseudocritical point scaling
changes to $L^{-2}(\ln{(\ln{L})})^{-1/2}$, from (\ref{FSSF}).

The $N$ independence of the leading multiplicative logarithmic corrections 
in the odd sector is fortuitous. That is the sector to which the rigorous 
results of \cite{AiGr83} concerning the triviality of the $O(N)$-$\phi^4_4$ theory
apply.  These results state that if the leading scaling behaviour of  the 
susceptibility differs from
that of mean field theory only by multiplicative  logarithmic corrections 
then the  theory must be trivial. The perturbative arguments above demonstrate that
the logarithmic corrections in the susceptibility are intimately linked to those
of the Lee-Yang zeroes and that these are, in fact, independent of $N$.
Thus independent numerical confirmation of their existence is strong evidence for
the triviality of the theory for all $N$. Such non-perturbative evidence
was provided in \cite{KeLa93,BuCo00,BiMa02} 
(see also \cite{ADCCaFr,RuGu85,LaMo90,AdSt97,N0logs}).

On the other hand, any future numerical attempts to detect the $N$ dependence 
of the leading logarithms must focus on the even sector. To our knowledge, the only
non-perturbative attempts that have so far been made to verify logarithmic corrections
in this sector are confined to $N=1$
\cite{RuGu85,LaMo90,KeLa93,AdSt97,BiMa02}.

%%%%%%%%%%%%%%%%%%%%%%%%%%%%%%%%%%%%%%%%%%%%%%%%%%%%%%%%%%%%%%%%%%%
\section{A Modified Finite Size Scaling Hypothesis}
%%%%%%%%%%%%%%%%%%%%%%%%%%%%%%%%%%%%%%%%%%%%%%%%%%%%%%%%%%%%%%%%%%%
\setcounter{equation}{0}
\label{sc:FSS}

The traditional FSS hypothesis for a thermodynamic function, $P$, is given in (\ref{FSShyp}). 
The hypothesis is based on the physical assumption that the only relevent length scales
involved are the actual length, $L$, of the finite-sized system and the correlation length,
$\xi_\infty (t)$,
of the bulk system. As a hypothesis, (\ref{FSShyp}) is a non-perturbative statement, but 
can be backed up by renormalization group arguments \cite{Br82}.

It has long been known that the standard FSS hypothesis, (\ref{FSShyp}), fails 
for $d\ge d_c$ \cite{LuBi99}. A modified hypothesis was proposed in \cite{KeLa93},
which correctly recovers the perturbative renormalization group predictions for 
FSS in the case of the single-component $\phi^4_4$-theory as well as recovering the 
traditional form of the hypothesis below four dimensions. 
In fact this modified hypothesis is sucessful for all $N$ in four dimensions, as 
demonstrated in the sequel (see also \cite{Ak01,BaFe98}).

In the modified FSS hypothesis, the actual length of the finite-sized system is replaced by
its correlation length, 
\begin{equation}
  \frac{P_L(t)}{P_\infty(t)}
  =
  {\cal{F}}_P\left(
   \frac{\xi_L(0)}{\xi_\infty (t)}
  \right)     
\quad .
\label{mFSShyp}
\end{equation}
The finite size behaviour of correlation length for four dimensions was calculated 
up to leading logarithms in \cite{Br82} as
\begin{equation}
 \xi_L(0) \propto
 L \left( \ln{L} \right)^\frac{1}{4}
\quad ,
\label{fssxi}
\end{equation}
independent of $N$.
A change in system size necessitates a
corresponding change in the temperature to keep the scaling variable, 
$x = \xi_L(0)/\xi_\infty (t)$, fixed. 
This change is
\begin{equation}
 |t| 
 \propto 
 x^2 L^{-2} \left( \ln{L} \right)^{\frac{N-4}{2(N+8)}}
\quad .
\label{switch}
\end{equation}
Inserting (\ref{switch}) into (\ref{two}) and (\ref{three})
recovers the perturbative renormalization group predictions
(\ref{fssforchiin4d}) and (\ref{FSSCv}) for the critical susceptibility
and specific heat, respectively.
The modified hypothesis, (\ref{mFSShyp}), 
also reduces to the traditional one, (\ref{FSShyp}), for $d< d_c$ as
$\xi_L(0) \propto L$ there.

Recently, Aktekin argued that the singular part of the free energy of the 
Ising model at the upper critical dimension obeys the Privman-Fisher-type ansatz 
\cite{Ak01,Lu02}
\begin{equation}
 f_L(t,H) = L^{-4} {\cal{F}}
 \left(
      tL^2\ln^{1/6}{L}, HL^3\ln^{1/4}{L}
 \right)
\quad .
\label{Aktekin}
\end{equation}
From this, the behaviour of the zeroes again follows. Setting $t=0$,
the partition function
takes the leading form of (\ref{tbi}) and hence  recovers the leading scaling for the 
Lee-Yang zeroes (\ref{FSSLY}).
 Likewise setting $h=0$ in (\ref{Aktekin}) 
recovers the correct form, (\ref{FSSF}), for the Fisher zeroes when $N=1$. 
A simple extension of Aktekin's formula, (\ref{Aktekin}), can now be proposed, which 
recovers the  FSS formulae in the general-$N$ case:
\begin{equation}
 f_L(t,h) = L^{-4} {\cal{F}}
 \left(
      tL^2(\ln{L})^{\frac{4-N}{2(N+8)}}, HL^3 (\ln{L})^{\frac{1}{4}}
 \right)
\quad .
\label{newak}
\end{equation}
As well as recovering (\ref{FSSLY}) in the odd sector, this  
recovers the  correct form, (\ref{FSSF}), for the 
zeroes in the even sector, except in the  $N=4$ case.
For the latter, the Aktekin's formula should be further modified, to read
\begin{equation}
 f_L(t,h) = L^{-4} {\cal{F}}
 \left(
      tL^2\left(\ln{\left({\ln{L}}\right)}\right)^{\frac{1}{2}}, HL^3
 (\ln{L})^{\frac{1}{4}}
 \right)
\quad .
\end{equation}
The functional form (\ref{newak}) also coincides with the large-$L$ behaviour of 
the free energy for $N\ne 4$ spin models with long-range interactions 
\cite{LuBl97}
(see also \cite{LuWi04,LuHe03}).

%%%%%%%%%%%%%%%%%%%%%%%%%%%%%%%%%%%%%%%%%%%%%%%%%%%%%%%%%%%%%%%%%%%
\section{Conclusions}
%%%%%%%%%%%%%%%%%%%%%%%%%%%%%%%%%%%%%%%%%%%%%%%%%%%%%%%%%%%%%%%%%%%
\setcounter{equation}{0}
\label{sc:ccl}

The issue of multiplicative 
logarithmic corrections to mean field scaling behaviour is one
of importance in both high energy and statistical physics. No rigorous
proof  of their presence in $O(N)$ $\phi^4_4$-theories exists, however 
there is strong analytical and numerical evedince that this is, in fact, 
the case.

Long before such logarithms were verified nonperturbatively in the Ising case 
\cite{KeLa93},
explicit calculations had been performed to see what subtleties simulators may 
expect in order to achieve a proper finite size extrapolation
there \cite{RuGu85}.
At present, there still exists no direct non-perturbative verification of the existence
and nature of these logarithms in the general $N$ case in four dimensions.
Here we have seen that any future numerical attempts to detect the 
$N$ dependence of the leading
logarithms will only come from the even sector. 
On the other hand, the fortuitous result that the logarithms in the odd sector 
are independent of $N$ means that a nonpertubative test of their behaviour in the 
Ising case may be interpreted as a test of triviality for all $N$ \cite{KeLa93}.

The results derived here are  in agreement with previous results in the literature, 
which  pertain specifically to the Ising ($N=1$ case)
and spherical  ($N \rightarrow \infty$ case) 
models at the upper critical dimension.

\bigskip
%
%%%%%%%%%%%%%%%%%%%%%%%%%%%%%%%%%%%%%%%%%%%%%%%%%%%%%%%%%%%%%%%%%%%

\end{document}